\documentclass[aps,prb,english,twocolumn,amsmath,showpacs]{revtex4}
\usepackage[latin1]{inputenc}
\usepackage{color}

\usepackage{graphicx}% Include figure files

\begin{document}

\title{Evidence of a {\sl s\/}-Wave Pairing in Heavily Zn-doped YBa$_{2}$Cu$_{3}$O$_{7-\delta}$ from Andreev Reflection Spectra}

\author{A. I.\ Akimenko\cite{AkinekoAI} and V. A.\ Gudimenko}

\affiliation{B.Verkin Institute for Low Temperature Physics and
Engineering National Academy of Sciences of Ukraine, 47 Lenin Ave.,
61103, Kharkiv, Ukraine}

\date{\today{}}

\begin{abstract}
When the {\it d$_{x^{2}-y^{2}}$}-wave pairing is suppressed by Zn-doping in YBa$_{2}$Cu$_{3}$O$_{7-\delta}$ some of the Andreev reflection spectra were found to be similar to the $s$-wave spectra of conventional superconductors. The energy gap is rather reproducible (2.3-3.0meV). It is suppressed by low magnetic field ($H_{C}^{PC}$=120-270mT) in great contrast to the $d$-wave spectra ($H_{C}^{PC}$$>$3T) with the similar order of the gap value. We suppose that the $s$-wave pairing occurs near the Zn-impurities.

\end{abstract}
\maketitle

The {\it d$_{x^{2}-y^{2}}$}-wave pairing is widely recognized as a dominant mechanism of superconductivity in the high temperature superconductors\cite{Tsuei}. However, in some cases an additional subdominant order parameter (OP) may better explain the experimental results\cite{Yeh,Daghero,Dagan,Kohen}. Theory predicts the appearance of {\it is}- or {\it id$_{xy}$}- subdominant OP near the (110)-surface where the {\it d$_{x^{2}-y^{2}}$}-wave OP is essentially suppressed due to the change of the order parameter sign along the quasiparticle trajectory\cite{Matsumota,Fogelstrom97,Tanuma}. The recent tunneling\cite{Akimenko06} and Andreev reflection\cite{Kohen} experiments are in agreement with the {\it is} subdominant OP in YBa$_{2}$Cu$_{3}$O$_{7-\delta}$ (YBCO). However, the problem is still under debates. We have applied the new approach here to clarify the question.

The order parameter may be also changed by doping. The Andreev refflection (AR) spectra show the possible transition from {\it d$_{x^{2}-y^{2}}$}- to {\it s}(or {\it d$\pm$is})-wave pairing with the oxygen doping change in Pr$_{2-x}$Ce$_{x}$CuO$_{4}$\cite{Biswas,Qazilbash}. It is well known that doping by Zn in YBCO decreases the critical temperature $T_C$ and energy gap\cite{Akimenko91,Akimenko95}. The critical temperature in YBa$_{2}$(Cu$_{1-x}$Zn$_{x}$)$_{3}$O$_{7-\delta}$ falls from about 90K to about 25K with the change of $x$ from zero to 0.075\cite{Roth}. That is why the heavily Zn-doped YBCO with low $T_C$ is perspective to find another pairing in YBCO and has been investigated here. We have found the typical for the {\it s}-wave superconductor Andreev refflection spectra that are very sensitive to low magnetic field in contrast to the {\it d}-wave spectra.

The first problem is how to distinguish the AR spectra (or the point-contact spectra when there is a barrier at interface) with the different type ({\it d$_{x^{2}-y^{2}}$}, {\it d$_{xy}$}, $s$) of OP.

%%%%%%%%%%%%%%%%%%%%%%%%%%%%%%%%%%%%%% FIG #1 %%%%%%%%%%%%%%%%%%%%%%%%%%%%%%%%%%%
\begin{figure}[b]
\begin{center}
\includegraphics[height=9.3cm, width=9cm,angle=0]{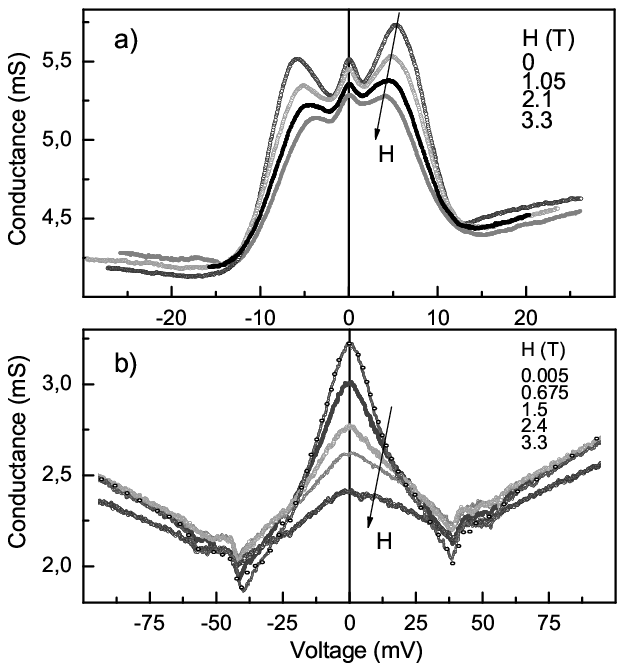}
\end{center}
\caption{Magnetic fields dependences of the {\it d$_{x^{2}-y^{2}}$}-wave Andreev reflection spectra of YBa$_{2}$(Cu$_{1-x}$Zn$_{x}$)$_{3}$O$_{7-\delta}$ with $x$=0.075.\\
a) shows the case of gap-related maximum presence (at $V\approx\pm$6mV for $H$=0). The position of the maximum in zero field may be different for different point-contacts (for instance, see the left inset in Fig.\,\ref{Unusual} and Ref.\,\onlinecite{Akimenko91}).\\
b) corresponds to the gapless superconductivity case.\\
The bath temperature $T$=4.2K.}\label{MagneticFields}
\end{figure}
%%%%%%%%%%%%%%%%%%%%%%%%%%%%%%%%%%%%%%%%%%%%%%%%%%%%%%%%%%%%%%%%%%%%%%%%%%%%%%%%%%

One of the evidence of the $d$-wave pairing is presence of the zero-bias conductance peak (ZBCP) in a tunneling spectrum (except the lobe-direction tunneling and gapless superconductor)\cite{Kashiwaya05}. For the point-contacts with direct conductivity, ZBCP must be absent if the barrier $Z$ at N/S boundary is zero. But due to the difference in Fermi velocity in the point-contact electrodes (a normal metal and high temperature superconductor) $Z$ is always more than zero\cite{Blonder83}, and ZPCB has been observed in most experiments\cite{Akimenko91,Akimenko95,Akimenko92,Goll,Akimenko94}. Boundary roughness, defects and impurities may decrease the intensity of ZBCP essentially\cite{Kashiwaya05}.

For the conventional $s$-wave superconductors, the modified BTK theory\cite{Plecenik} discribes an experimental point-contact (PC) spectrum quite well using three fitting parameters: a gap value $\Delta$, $Z$ and relatively small smearing factor $\Gamma$, $\Gamma$/$\Delta$$\ll$1. In the case of the $d$-wave superconductor it is not usually possible to do that with the reasonable value of $\Gamma$. It is because of the strong gap anisotropy and low angle resolution for the orifice-like point-contacts\cite{Kulik} ($\sim$90°). However, the channel-like point-contacts with rough walls may have the angle resolution close to a tunnel junction. Thus, even for the $d$-wave superconductor, it is possible to register a PC spectrum which is similar to the $s$-wave one if, for instance, the channel-like point-contact in [100] (or [010]) direction is realized in our experiment. In different point-contacts the direction of electron flow is different as a rule, and the gap values extracted may be essentially different for the same $T_C$ in the point-contact region. Taking into account the possible complex configuration of a real point-contact (several conducting spots with different shape), analysis only of the PC spectrum form is not reliable method to know the OP type. Nevertheless, the information about gap distribution for the $d$-wave superconductor may be obtained by the proper histogram building\cite{Akimenko91}.

The critical parameters ($T_C$ and $H_{C2}$) of the {\it d}-wave superconductor was found to be much higher than that for the conventional {\it s}-wave one (for YBa$_{2}$Cu$_{3}$O$_{7-\delta}$ $T_C$$\approx$95K and $H_{C2}$$>$100T )\cite{Smith}.

%%%%%%%%%%%%%%%%%%%%%%%%%%%%%%%%%%%%%% FIG #2 %%%%%%%%%%%%%%%%%%%%%%%%%%%%%%%%%%%
\begin{figure}[b]
\begin{center}
\includegraphics[width=9cm,angle=0]{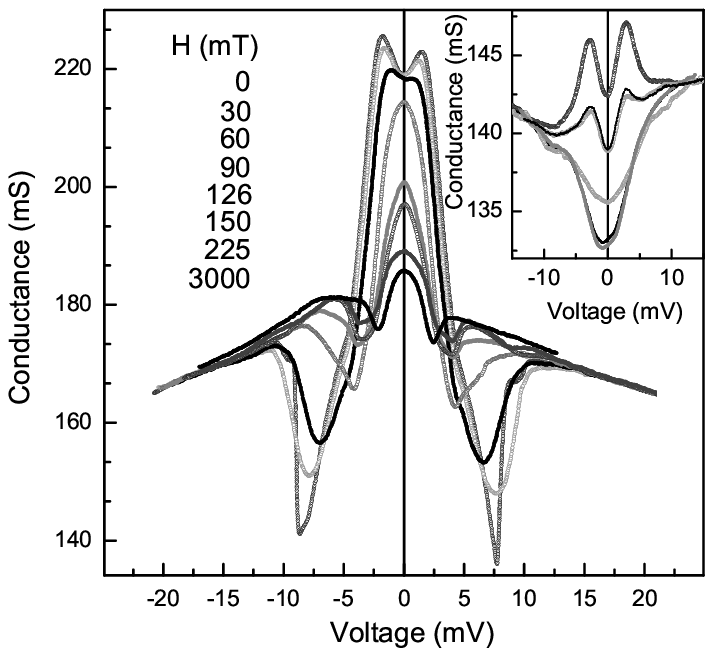}
\end{center}
\caption{The {\it s}-wave type Andreev reflection spectra of YBa$_{2}$(Cu$_{1-x}$Zn$_{x}$)$_{3}$O$_{7-\delta}$ ($x$=0.075) with the magnetic field change. $T$=4.2K. $H_{C}^{PC}\approx$225mT.\\
Inset shows the case with another character of the background behavior at $H\ge H_{C}^{PC}\approx$120mT. H=0, 30, 60, 120, 225, 450, 600mT from upper curve.}\label{sWaveType}
\end{figure}
%%%%%%%%%%%%%%%%%%%%%%%%%%%%%%%%%%%%%%%%%%%%%%%%%%%%%%%%%%%%%%%%%%%%%%%%%%%%%%%%%%

{\it Thus, if one will register the PC spectra without ZBCP with low critical parameters and similar gap values for different PCs, pairing is very possible to be of the $s$-wave type.}

It was earlier found that doping by Zn decreases $T_C$ without essential change of electron density in YBa$_{2}$(Cu$_{1-x}$Zn$_{x}$)$_{3}$O$_{7-\delta}$\cite{Kawaji,Li}. At {\it x}$\approx$0.1-0.12, $T_C$ falls to about 10K. The distribution of energy gaps also goes to zero\cite{Akimenko91}. At {\it x}$\ge$0.05, some of the PC spectra look like expected for the gapless superconductor\cite{Akimenko93}. Most likely, the gapless state appears on the part of Fermi surface close to the node lines.

We have investigated the polycrystal sample with the nominal {\it x}=0.075. The resistivity measurement shows the wide transition from normal to the superconducting state (from 40K to 10K). It is in agreement with $T_C$$\approx$25K for {\it x}=0.075 obtained in the sample with the steeper resistive transition in Ref.\,\onlinecite{Akimenko91}. High inhomogeneity in the Zn distribution let us getting the large variety of the PC spectra in the same experiment to study the magnetic field effect.

The standard modulation method\cite{Blackford} was applied to  measure {\it dI/dV} vs $V$.

%%%%%%%%%%%%%%%%%%%%%%%%%%%%%%%%%%%%%% FIG #3 %%%%%%%%%%%%%%%%%%%%%%%%%%%%%%%%%%%
\begin{figure}[t]
\begin{center}
\includegraphics[width=9.2cm,angle=0]{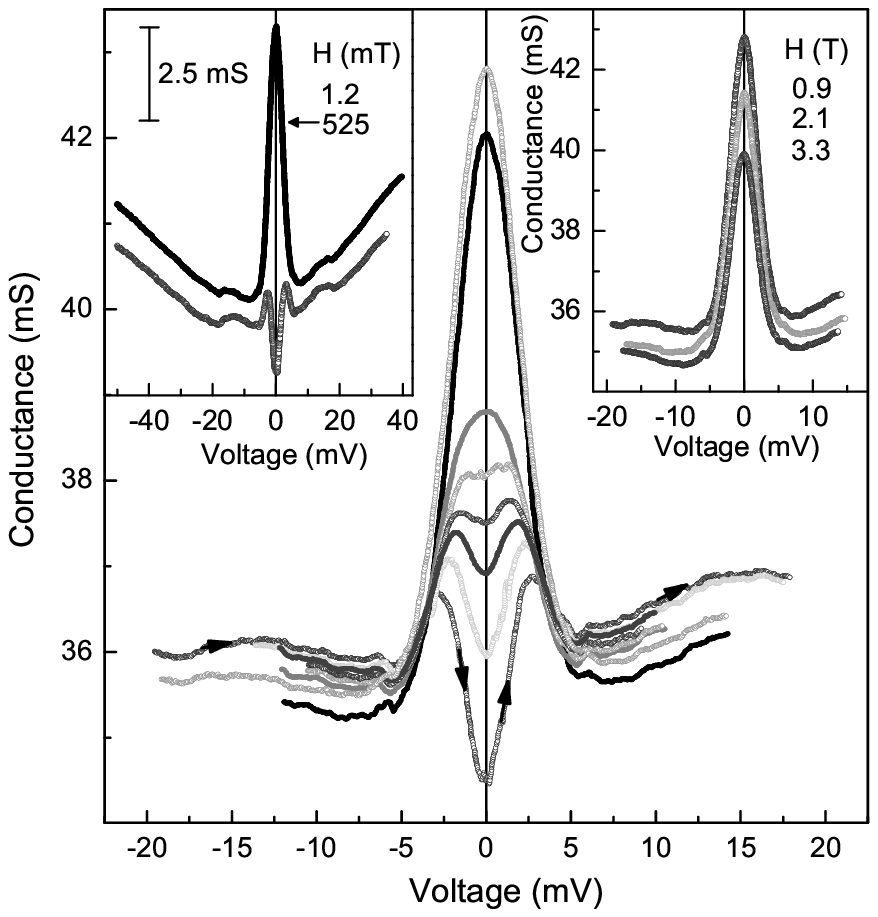}
\end{center}
\caption{Unusual transformation of the s-wave type spectrum with the magnetic field. {\it H}=1.2, 105, 150, 180, 225, 270, 600, 900mT starting from the bottom at {\it V}=0. $T$=4.2K. $H_{C}^{PC}\approx$270mT.\\
Right inset shows the high magnetic field effect.\\
Left inset shows two spectra in the enlarged bias range. The spectrum at {\it H}=525mT is shifted up for clarity.}\label{Unusual}
\end{figure}
%%%%%%%%%%%%%%%%%%%%%%%%%%%%%%%%%%%%%%%%%%%%%%%%%%%%%%%%%%%%%%%%%%%%%%%%%%%%%%%%%%

%%%%%%%%%%%%%%%%%%%%%%%%%%%%%%%%%%%%%% FIG #4 %%%%%%%%%%%%%%%%%%%%%%%%%%%%%%%%%%%
\begin{figure*}[t]
\begin{center}
\includegraphics[width=18cm,angle=0]{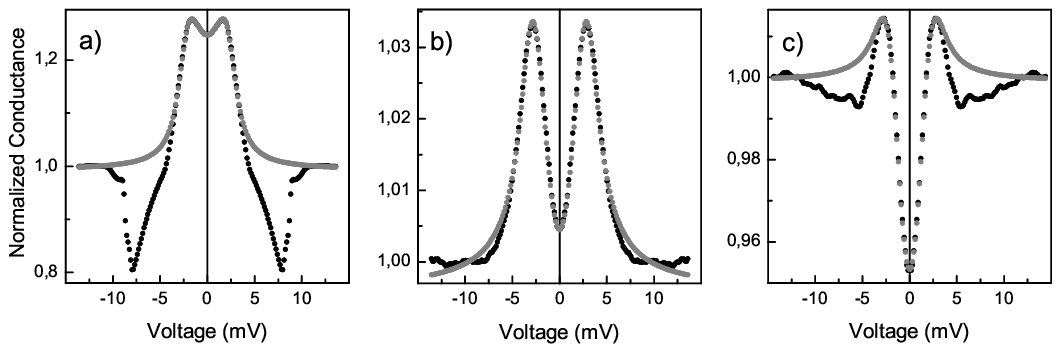}
\end{center}
\caption{Comparison of the experimental s-wave spectra (dark dots) with the BTK calculations. The experimental curves were previously symmetrized and normalized on value at $V$=$\pm$15mV. The parameters of fitting are as follows:\\
a) $\Delta$=2.35meV, $\Gamma$=0.2meV, $Z$=0.30
\ \ \ b) $\Delta$=3.00meV, $\Gamma$=0.4meV, $Z$=0.55
\ \ \ c) $\Delta$=2.40meV, $\Gamma$=0.4meV, $Z$=1.15}\label{Comparison}
\end{figure*}
%%%%%%%%%%%%%%%%%%%%%%%%%%%%%%%%%%%%%%%%%%%%%%%%%%%%%%%%%%%%%%%%%%%%%%%%%%%%%%%%%%

In Fig.\,\ref{MagneticFields}\,, two kinds of the PC spectra (with low $Z$) typical for the heavily Zn-doped YBCO ({\it x}$\ge$0.05) are shown. The first (a) has the gap-related maximum and relatively narrow ZBCP, the second (b) has a wide maximum around $V$=0. Theory predicts approximately such a form of PC spectrum for the gap- and gapless- superconductor respectively\cite{Beloborodko91,Beloborodko03}.

The magnetic field of about 3$T$ affects both observed peaks essentially while in the Zn-undoped YBCO, such a field has no any visible effect on our PC spectra (except the ZBCP). In the case (a), the magnetic field shifts the gap-related maximum at $V\approx\pm$6mV to lower energies like it was found earlier for conventional superconductors\cite{Naidyuk96,Miyoshi}. The absence of splitting of ZBCP with the field was observed earlier in the tunneling and point-contact experiments too\cite{Qazilbash,Ekin,Alff}, and one of the possible reason is that the field is parallel to the N/S interface\cite{Aprili,Aubin}. Our point-contacts were made between the rod-shape electrodes (like in Ref.\,\onlinecite{Chubov}), and geometrically, the magnetic field was applied parallel to the N/S interface. However, the real situation is difficult to control because of surface roughness.

In the case of gapless regime (b), the field suppresses the peak around $V$=0 without any essential change of its energy location and form. Such a behavior is well known for the conventional gapless superconductors in tunneling experiments\cite{NewYork69}.

Thus, it seems that there is no any difference in magnetic field effect (except the value applied) on the {\it d$_{x^{2}-y^{2}}$}-wave superconductor YBCO and the conventional {\it s}-wave superconductor assuming that Zn-doping does not change the pairing mechanism.

We have also registered some spectra (Fig.\,\ref{sWaveType}\, and \,\ref{Unusual}) similar to found ones in the numerous studies of the conventional {\it s}-wave superconductors. They have clear maximum at low energy without any ZBCP. In Fig.\,\ref{Comparison}\,, the symmetrized experimental curves measured at $H\approx$0 are compared with the calculated ones. There is a good agreement in the gap-related region (interval about $\pm$5mV around zero bias). The structure at $\approx$5-10mV seen on the curve a) and c) is often observed, but its origin is not clear yet\cite{Sheet,Shan,Strijkers}. The modified BTK fitting procedure\cite{Plecenik} gives the similar values of gap $\Delta$=2.35-3.0meV for different point-contacts with the small enough smearing factor $\Gamma$/$\Delta<$0.2. It is necessary to note that the gap value extracted from the point-contact experiment may be different if the bulk gap value depend on pressure. The pressure in the mechanically made point-contacts may be rather different, and it may be a reason of the gap value variation found.

The most interesting finding is that all the spectra are very sensitive to low magnetic field in great contrast to the shown in Fig.\,\ref{MagneticFields}\,. The gap-related maximum goes to zero bias in a way characteristic to the conventional $s$-wave superconductor\cite{Naidyuk96,Miyoshi}. It is most clear seen for curves in Fig.\,\ref{sWaveType}\,. The critical magnetic field for the point-contact region $H_{C}^{PC}$ corresponds to the case when the AR spectrum is entirely suppressed, and for the PC spectra of the $s$-wave type registered here is 120-270mT.
The spectrum in Fig.\,\ref{Unusual}\, demonstrates more complex shape and behavior with the increasing magnetic field. It has also the weaker structure at $V$=$\pm$10-20mV that is not so changeable with magnetic field (see left inset in Fig.\,\ref{Unusual}) as the low-bias structure (at $V\le$5mV). This high bias structure similar to that found in YBa$_{2}$(Cu$_{1-x}$Zn$_{x}$)$_{3}$O$_{7-\delta}$ (with $x$=0.025) gap-related structure\cite{Akimenko91}. The low-bias double-peak structure transforms into the high ZBCP that is suppressed in rather high magnetic field in a way similar to the spectrum shown in Fig.\,\ref{MagneticFields}b\, (gapless regime). This observation says that there are different superconducting phases near N/S boundary in this point contact. The different reaction on the magnetic field applied may give such an unusual behavior.

We have also noticed that the $s$-type spectrum is usually registered after the spectrum with the resonance scattering peak similar to that found in Refs.\,\onlinecite{Pan} and \,\onlinecite{Yeh}. Because our point-contacts were made by the successive mutual shift of electrodes\cite{Chubov}, one can suppose that the $s$-type superconductivity locates near a Zn-impurity (or a cluster).

Proximity effect, induced by the {\it d$_{x^{2}-y^{2}}$}-superconductor, may be a reason of the $s$-wave pairing\cite{Ohashi} in a normal electrode, but in Andreev reflection effect the conductance maximum (or a kink) is connected only with the maximum gap along the quasiparticle trajectory\cite{Blonder82}. The "proximity gap" is always less than in {\it d$_{x^{2}-y^{2}}$}-superconductor.

In summary, we have found that some of the Andreev-reflection spectra observed in the heavily Zn-doped YBa$_{2}$Cu$_{3}$O$_{7-\delta}$ are similar to the conventional $s$-wave superconductor ones in a shape, gap value reproducibility and sensitivity to low magnetic field. It confirms the appearance of the $s$-wave pairing in YBa$_{2}$Cu$_{3}$O$_{7-\delta}$ if the $d$-wave pairing is suppressed.

We acknowledge Prof. I. K. Yanson for the useful discussions and Prof. S. Suleimanov for the YBCO samples.

\end{document}